\documentclass[aps,floatfix,nofootinbib]{revtex4}
\usepackage{graphics}
\usepackage{graphicx}
\usepackage{amssymb}		
\usepackage{amsmath}
\usepackage{verbatim}
\usepackage{axodraw}
\usepackage{pstricks}
\usepackage{slashed}

\newcommand{\gsim}{\!\mathrel{\hbox{\rlap{\lower.55ex \hbox{$\sim$}} \kern-.34em \raise.4ex \hbox{$>$}}}}
\begin{document}

\setlength{\baselineskip}{0.22in}
\begin{flushright}MCTP-07-16\\
\end{flushright}
\vspace{0.2cm}

\title{Sfermion Interference in Neutralino Decays at the LHC}

\author{Daniel J. Phalen and Aaron Pierce}
\vspace{0.2cm}
\affiliation{Michigan Center for Theoretical Physics (MCTP) \\
Department of Physics, University of Michigan, Ann Arbor, MI
48109}

\date{\today}

\begin{abstract}
If the two lightest neutralinos of the Minimal Supersymmetric Standard Model have a mass splitting less than the $Z$ boson mass, interference effects in the three-body decay $\tilde{\chi}_2^0 \rightarrow \tilde{\chi}_1^0 f \overline{f}$  can be important.  We formulate an observable that contains information on the nature of the interference: the ratio $\rm{BR}(\tilde{\chi}_2^0 \rightarrow \tilde{\chi}_1^0 b \overline{b})/\rm{BR}(\tilde{\chi}_2^0 \rightarrow \tilde{\chi}_1^0 l^+ l^-)$.   This will give a constraint on the supersymmetry breaking parameters that is complementary to many techniques already existing in the literature.  We present some ideas on how to perform a simple counting experiment to determine this ratio. 
\end{abstract}

\maketitle

\setcounter{equation}{0}

%%%%%%%%%%Introduction
\section{Introduction}
Supersymmetry (SUSY) offers one of the best theoretically motivated models for new physics at the weak scale.  If the sparticles are not too heavy, the Large Hadron Collider (LHC) will copiously produce them, leading to striking signatures that should be easy to observe above Standard Model backgrounds.  It will be harder to find the masses and couplings of the superpartners. Several promising methods exist using cascades of colored superpartners, see, e.g., \cite{ATLAStdr, Hinchliffe:1996}.

Because of their small direct production cross section, measurements of uncolored superpartners, e.g. sleptons, are challenging if they do not appear on-shell in the decay chain of the colored particles.  It is possible, however, to see the effects of these particles indirectly.  If $\tilde{\chi}_2^0$ undergoes a three-body decay to $\tilde{\chi}_1^0 f \overline{f}$,  the corresponding sfermion, $\tilde{f}$, while not appearing on-shell, still affects the branching ratios and invariant mass distribution of the fermions in this decay. This observation has been pointed out as a way to study sleptons in~\cite{Nojiri:1999} and~\cite{Birkedal:2005}.  Both of these references presuppose large statistics to make precise measurements of invariant mass distribution of lepton pairs. In the case of large $\tan{\beta}$, similar discussions were made regarding the stau in \cite{Baer:1998}.

Extending these works, we propose looking at the ratio of  $\tilde{\chi}_2^0 \rightarrow \tilde{\chi}_1^0 b \overline{b}$ to $\tilde{\chi}_2^0 \rightarrow \tilde{\chi}_1^0 l^+ l^-$ events in gluino cascade decays. This ratio will be independent of the gluino mass and branching fractions, and should provide some utility even with a relatively small number of events. This sample should be relatively easy to isolate above Standard Model backgrounds due to the presence of multiple energetic jets and leptons paired with large amounts of missing energy.  We will show that a measurement of this ratio is a goal that is both worthwhile and likely attainable in the case where there is a light gluino and the mass splitting between the two lightest neutralinos is less than $m_{Z}$.

At large values of the slepton and b-squark masses this ratio will be independent of the parameters of the neutralino mass mixing matrix, and will be solely governed by couplings to the $Z$ boson.  At lower sfermion masses the sfermion exchange diagrams become important and could alter this ratio.  For example, there is destructive interference of the slepton exchange diagram and Z-exchange diagrams for light sleptons~\cite{Baer:1995,Mrenna:1996}.  The ratio can give important information about the superpartner spectrum.

The outline of this paper is as follows.  In section 2 we introduce our proposed measurement and discuss the interference in neutralino decays in some detail.  In section 3 we examine how our proposed measurement complements a measurement of the shape of the lepton invariant mass distribution in neutralino decays.  In section 4 we look at the signatures at the LHC for a specific spectrum and how these signatures would be affected by the mass of the b-squark.  In section 5 we talk about deformations of the superpartner spectrum and their impact on our measurement. 
Finally, we draw some conclusions.

%%%%%%%%%%Proposed Measurement
\section{Proposed Measurement}
Since we would like to measure 3-body decays $\tilde{\chi}_2^0 \rightarrow \tilde{\chi}_1^0 f \overline{f}$, we start by requiring that $m_{\tilde{\chi}_2^0} - m_{\tilde{\chi}_1^0} < m_Z$.  This prevents on-shell $Z$ decays, and three-body decays mediated by sfermions, $Z$ bosons, and Higgs bosons can all be competitive. Consider the quantity
\begin{equation}
R \equiv \frac{\#(\tilde{\chi}_2^0 \rightarrow \tilde{\chi} b \overline{b})}{\#(\tilde{\chi}_2^0 \rightarrow \tilde{\chi} l \overline{l})} = \frac{BR(\tilde{\chi}_2^0 \rightarrow \tilde{\chi} b \overline{b})}{BR(\tilde{\chi}_2^0 \rightarrow \tilde{\chi} l \overline{l})} =  \frac{\Gamma(\tilde{\chi}_2^0 \rightarrow \tilde{\chi} b \overline{b})}{\Gamma(\tilde{\chi}_2^0 \rightarrow \tilde{\chi} l \overline{l})},
\end{equation}
where $l$ is taken to be electrons and muons\footnote{In the rest of the paper, when we say leptonic decays we will be limiting ourselves to $\mu$ and $e$ as $l$.  One can also measure a quantity analogous to $R$ using $\tau$'s with same method described in our paper and some of the techniques described by Baer et al. in~\cite{Baer:1998}.  In that paper they also mention measuring $R_{\tau} = BR(\tilde{\chi}_2^0 \rightarrow \tilde{\chi}_1^0 \tau^+ \tau^-)/  BR(\tilde{\chi}_2^0 \rightarrow \tilde{\chi}_1^0 e^+ e^-)$ or with $\mu$ instead of $e$, this test of slepton (non)-universality is particularly important at large $\tan{\beta}$. }.  If we can isolate these $\tilde{\chi}_2^0$ decays to $\tilde{\chi}_1^0$ (an assumption which we discuss later) , $R$ eliminates the dependence on the gluino production cross section and branching ratios.  

\begin{figure}
  \begin{center}
    \scalebox{1.5}{\includegraphics{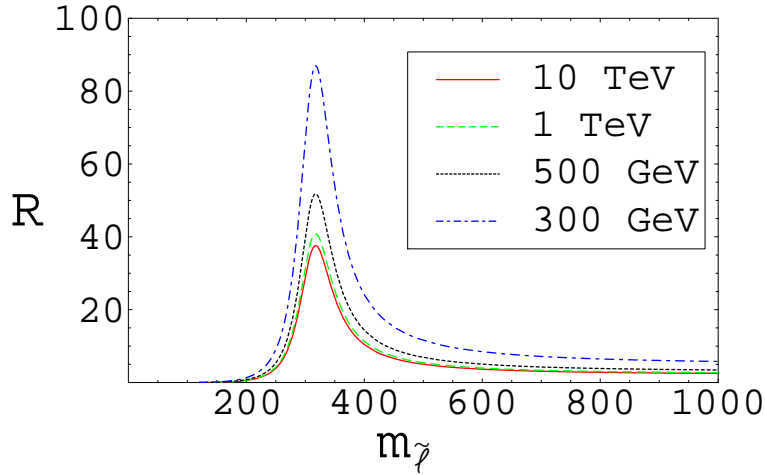}}
  \end{center}
    \caption{Here we plot $R$ as a function of slepton mass for different values of the b-squark mass.  The plateau on the right for $m_{\tilde{b}} = 1$ TeV is the decoupling limit for sleptons and b-squarks.  Here we use $M_1 = 70$ GeV, $M_2=140$ GeV, $\mu=300$ GeV, and $\tan{\beta}=4$. Left- and right-handed sfermion masses are assumed to be degenerate.
  \label{fig:Ratiopt1graphs}}
\end{figure}

When the slepton mediated, b-squark mediated, and Higgs mediated diagrams decouple, $R$ is completely determined by the coupling of the $Z$ to the fermions.  In this decoupling limit, $R = R_{z} \approx 2.2$ for $m_b << m_{\tilde{\chi}_2^0} - m_{\tilde{\chi}_1^0}$.  Adding in fermion masses will introduce a small dependence on the mass difference of the two neutralinos and pushes $R$ slightly lower.  In Figure~\ref{fig:Ratiopt1graphs}, $R$ is plotted as a function of slepton mass for some different values of the b-squark mass.  For the neutralino masses chosen in Figure~\ref{fig:Ratiopt1graphs}, $m_b << M_2 - M_1 < m_Z$ and $R_{z} \approx 2.1$.   In that Figure the domain where the $Z$-exchange dominates can be seen as the plateau at large sfermion masses\footnote{Note that our choice of $M_{2}$ and $M_{1}$ corresponds roughly to unified boundary conditions at the Grand Unified scale.  We do not impose a GUT relation on the gluino mass.  If the gluino mass is too heavy, then there may be too few events for an effective measurement of R.}.

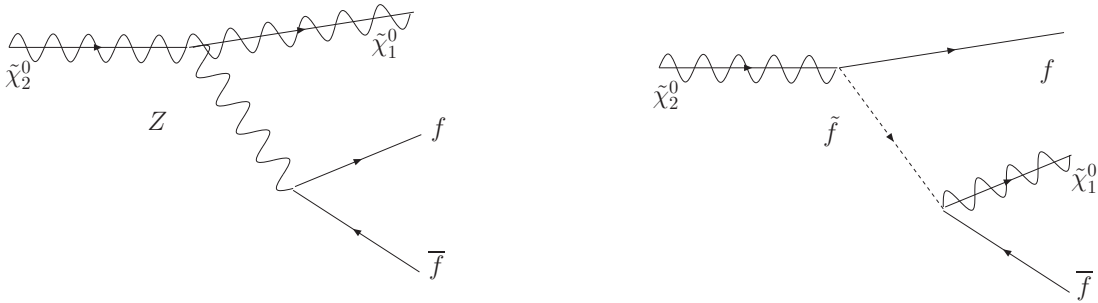
\begin{figure}
\begin{center}
  \scalebox{0.7}{
  \begin{picture}(620,150) (7,0)
    \SetWidth{0.5}
    \SetColor{Black}
    \ArrowLine(9,125)(105,125)
    \ArrowLine(106,125)(227,144)
    \Photon(9,125)(104,124){7.5}{5}
    \Photon(104,124)(225,143){7.5}{6}
    \Photon(108,125)(162,49){7.5}{5}
    \ArrowLine(163,50)(231,78)
    \Text(84,81)[lb]{\Large{\Black{$Z$}}}
    \ArrowLine(230,4)(162,48)
    \Text(7,103)[lb]{\Large{\Black{$\tilde{\chi}_2^0$}}}
    \Text(205,120)[lb]{\Large{\Black{$\tilde{\chi}_1^0$}}}
    \Text(237,75)[lb]{\Large{\Black{$f$}}}
    \Text(236,0)[lb]{\Large{\Black{$\overline{f}$}}}
    \ArrowLine(359,114)(455,114)
    \ArrowLine(456,114)(577,133)
    \Photon(359,114)(454,113){7.5}{5}
    \ArrowLine(513,39)(581,67)
    \ArrowLine(580,-7)(512,37)
    \Text(357,92)[lb]{\Large{\Black{$\tilde{\chi}_2^0$}}}
    \Text(586,-11)[lb]{\Large{\Black{$\overline{f}$}}}
    \DashArrowLine(456,114)(512,38){2}
    \Text(566,106)[lb]{\Large{\Black{$f$}}}
    \Text(583,47)[lb]{\Large{\Black{$\tilde{\chi}_1^0$}}}
    \Photon(512,39)(580,67){7.5}{4}
    \Text(449,72)[lb]{\Large{\Black{$\tilde{f}$}}}
  \end{picture}
  }
\end{center}
\caption{The two diagrams that interfere to give the $\tilde{\chi}_2^0 \rightarrow \tilde{\chi}_1^0 f \overline{f}$ decay amplitude. For a calculation of the partial width, see Appendix A. \label{fig:neutrfeyn}}
\end{figure}

The calculation of the three body decay rate $\tilde{\chi}_2^0 \rightarrow \tilde{\chi}_1^0 f \overline{f}$ is outlined in the Appendix.  Contributions to the process can be classified as $A_{IJ}$, where $I$ labels the intermediate sfermion 
and $J$ is the helicity of the final state fermions.  The relevant diagrams are shown in Figure~\ref{fig:neutrfeyn}.  The decay is often dominated by $A_{LL}$, the amplitude coming from the case where both fermion and sfermion are left handed:
\begin{equation}
  A_{LL} = \frac{g_Z^2}{2} \frac{ z_{BA}^{\tilde{\chi}^0} (T_3 - Q_f \sin^2{\theta_W}) }{(q+\bar{q})^2-m_Z^2} - \frac{g^2}{2} \frac{a_{AL}^f a_{BL}^f}{(\bar{p}+\bar{q})^2-m^2_{\tilde{f}_L}}- \frac{g^2}{2} \frac{a_{AR}^f a_{BR}^f}{(\bar{p}+\bar{q})^2-m^2_{\tilde{f}_R}}. \label{eq:Smatrix}
\end{equation}
Here $g_Z=\sqrt{g^2 + g^{'2}}$ is the $Z$ gauge coupling; $g$ is the $SU(2)_{L}$ gauge coupling; $z_{BA}^{\tilde{\chi}^0}$ is the $Z$ coupling to neutralinos; $T_3$ is the weak isospin quantum number; $Q_f$ is the charge of the fermion, and $a^f_{AX}$ is the $\tilde{\chi}_A^0 \tilde{f}_L f_X$ coupling.  We do not include the Higgs boson exchange diagrams.  The Higgs boson exchange is important in the case where $\tan{\beta} \gsim 45$ or $m_{A^0}$ is small, resulting in large $\tau$ and b-quark Yukawa couplings or larger values of the $Z \tilde{\chi}_2^0 \tilde{\chi}_1^0$ coupling.  This case is studied in~\cite{Baer:1998} and \cite{Bartl:1999}.  Here we limit ourselves to the diagrams shown in Figure~\ref{fig:neutrfeyn}, a good approximation in the limit where the Higgs boson contributions are small. 

This dominant contribution will determine whether the sfermion destructively or constructively interferes.  We would like to analyze in what cases the two diagrams in Figure~\ref{fig:neutrfeyn} will destructively interfere, which will be a function of the quantum numbers of the exchanged sfermion, $\tilde{f}$.  First, consider $Z$ boson exchange.  The $Z$ diagram has a coupling proportional to $z_{BA}^{\tilde{\chi}^0} (T_3 -Q_f \sin^2{\theta_W})$ for decays to left-handed fermions.  In the limit $\mu >> m_{Z}$, one can perturbatively diagonalize the neutralino mass matrix in the spirit of \cite{Arnowitt:1995}, finding: 
\begin{equation}
z_{12}^{\tilde{\chi}^0} \varpropto \left(1 + \frac{M_1-M_2}{2\mu} -\frac{(M_1+M_2)M_1}{2\mu^2}\right)\cos{2\beta}. 
\end{equation}
So for $\tan{\beta} > 1$ and $M_1 , M_2 << \mu$ we would expect that $z^{\tilde{\chi}^0}_{12} < 0$.  Then for left-handed down-type fermions (where $T_3 = -1/2$), the combination $z_{BA}^{\tilde{\chi}^0} (T_3 -Q_f \sin^2{\theta_W})$ is positive.  Now we turn to sfermion exchange.  Looking at the couplings in the sfermion diagrams in the limit of $\mu >> m_Z$ gives an expression for the coupling to the left-handed fermions
\begin{eqnarray}
  a^f_{1L}  a^f_{2L} \sim 4 T_3 Y_L \tan{\theta_W},
\end{eqnarray}
where $Y_L$ is the hypercharge of left handed fermions.  For the b-squark this is negative, while for the sleptons this is positive.  Since the first and second terms in Eqn.~(\ref{eq:Smatrix}) appear with opposite sign, there is constructive interference for decays to b-quarks and destructive interference for decays to leptons.  Therefore, in the limit $\mu >> m_Z$ there is substantial destructive interference from the left-handed sleptons for some mass range.  This is the source of the bump in Figure~\ref{fig:Ratiopt1graphs}.  A similar analysis for the case of right-handed sleptons finds constructive interference.  With the assumption of equal masses for the right-handed and left-handed sleptons, this interference is numerically sub-dominant.  We assume this degeneracy in masses unless otherwise stated.  In the next section, we will outline a simple counting experiment that is sensitive to the interference described here.

As long as the sfermion exchange diagram has not decoupled, we potentially can gain information about the virtual sfermion.  The decoupling depends not only on the masses of the sfermions, but also the detailed structure of the neutralino mass matrix, which in turn determines the couplings in both the sfermion and $Z$ exchange diagrams.  The $Z \tilde{\chi}_2^0 \tilde{\chi}_1^0$ coupling, $z_{12}^{\tilde{\chi_{0}}}$, is mainly governed by $\mu$ because the $Z$ couples to the Higgsino components of the neutralinos.  A higher value of $|\mu|$ implies less gaugino mixing with Higgsinos, suppressing the $Z$ exchange diagram.  This, in turn,  means that sfermion diagrams affect the partial width for higher values of the sfermion mass, and so for larger $|\mu|$ (say, 800 GeV) the decoupling limit may not be fully reached until both the slepton and b-squark masses are 10 TeV or higher. 
We determine the combination $a_{2L}^f a_{1L}^f$ makes the largest contribution to the sfermion mediated diagrams, largely due to the wino content present in the two lightest neutralinos.  Then we can identify 
\begin{equation}
m_{\tilde{f}_L}^2 >> \frac{g^2}{g_Z^2}\frac{a_{2L}^f a_{1L}^f}{z_{12}^{\tilde{\chi}^0} (T_3 - Q_f \sin^2{\theta_W})}m_Z^2
\end{equation}
as the decoupling limit. 
For sfermion masses that do not satisfy the above criterion, there will be significant modification of $R$, and thus it is possible to gain indirect information about them.

If $R$ is found to deviate from $R_{z}$, either b-squark or slepton interference is important. The direction of the deviation can give an important clue as to what sfermions are interfering.  If $R < R_z$ the slepton diagram is dominating the neutralino decay to leptons.  For $R > R_z$, either sleptons are interfering destructively or b-squarks are interfering constructively.   

%%%%%%%Things that R tells us
\section{$R$ and the lepton invariant mass distribution}
In this section we emphasize the complementary nature of $R$ and the lepton invariant mass distribution.  Neither contains complete information.  $R$ is sensitive to the the partial width $\Gamma(\tilde{\chi}_2^0 \to \tilde{\chi}_1^0 l^+ l^-)$, while the invariant mass shape is 
\begin{equation}
\frac{1}{\Gamma(\tilde{\chi}_2^0 \to \tilde{\chi}_1^0 l^+ l^-)} \frac{d\Gamma(\tilde{\chi}_2^0 \to \tilde{\chi}_1^0 l^+ l^-) }{ d m_{ll}}.
\end{equation}

\begin{figure}
  \begin{center}
    \scalebox{1.5}{\includegraphics{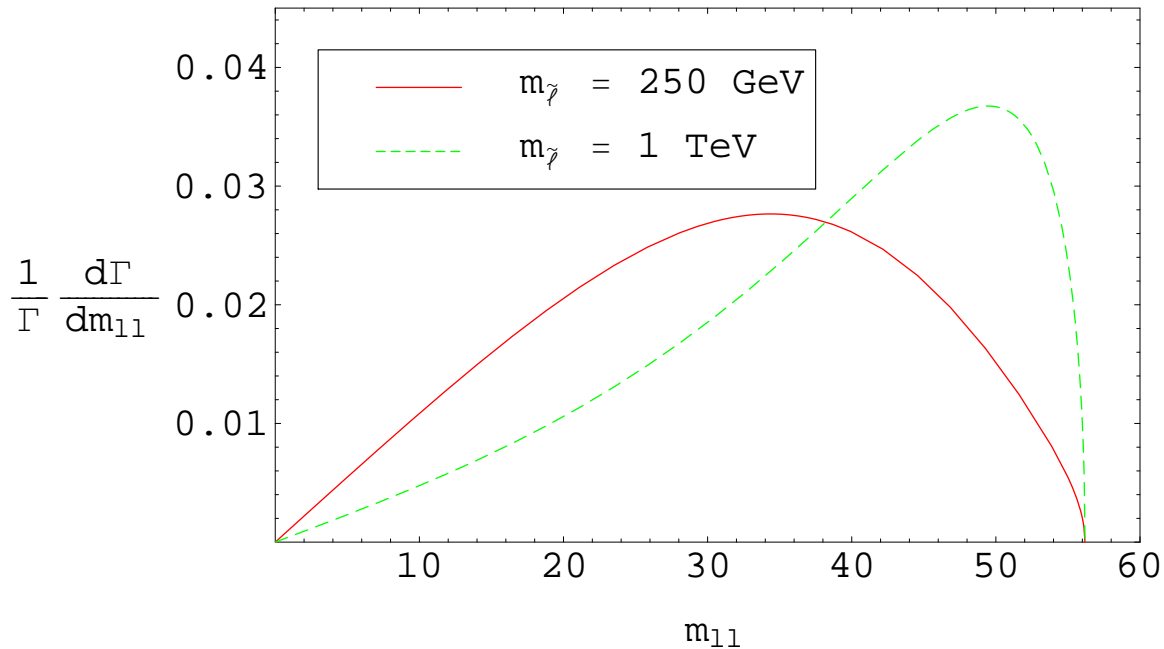}}
  \end{center}
    \caption{Here we plot lepton invariant mass distributions for $R=R_z$ for the decoupling limit and for the special value of the slepton mass corresponding to the left side of the bump in Figure~\ref{fig:Ratiopt1graphs}.  In this figure, $M_1 = 70$ GeV, $M_2=140$ GeV, $\mu=300$ GeV, and $\tan{\beta}=4$. \label{fig:Rdegen}}
\end{figure}

To see that the invariant mass distribution can complement R, note that there is a degeneracy between points on opposite sides of the bump in Figure~\ref{fig:Ratiopt1graphs}.   There is even a case when $R = R_{z}$ at low slepton mass -- even though the slepton interference is important, the $Z$ and slepton mediated diagrams can conspire to recover $R = R_{z}$.  Fortunately, we can lift this degeneracy even in the case of low statistics by examining the invariant mass distribution of the two leptons, a la \cite{Birkedal:2005,Nojiri:1999}.  In Figure~\ref{fig:Rdegen}, we show invariant mass distributions for two different slepton masses, 1 TeV and 250 GeV that have similar values for $R$ (on opposite sides of the bump in Figure~\ref{fig:Ratiopt1graphs}).  For a 1 TeV slepton mass $Z$ exchange dominates the neutralino decays. In this case, the invariant mass of the leptons is pushed toward $m_Z$ to maximize the momentum flow through the virtual $Z$.  For the 250 GeV slepton mass, the invariant mass distribution of the leptons is peaked in the center.  This is because the momentum flow through the virtual slepton maximizes the invariant mass of $\tilde{\chi}_1^0$ and one lepton.  Even a very coarse binning procedure would differentiate these two cases.  Some powerful techniques for differentiating invariant mass distributions in the case of higher statistics are outlined in~\cite{Birkedal:2005} using the Kolmogorov-Smirnov test.  

\begin{figure}
  \begin{center}
    \scalebox{1.2}{\includegraphics{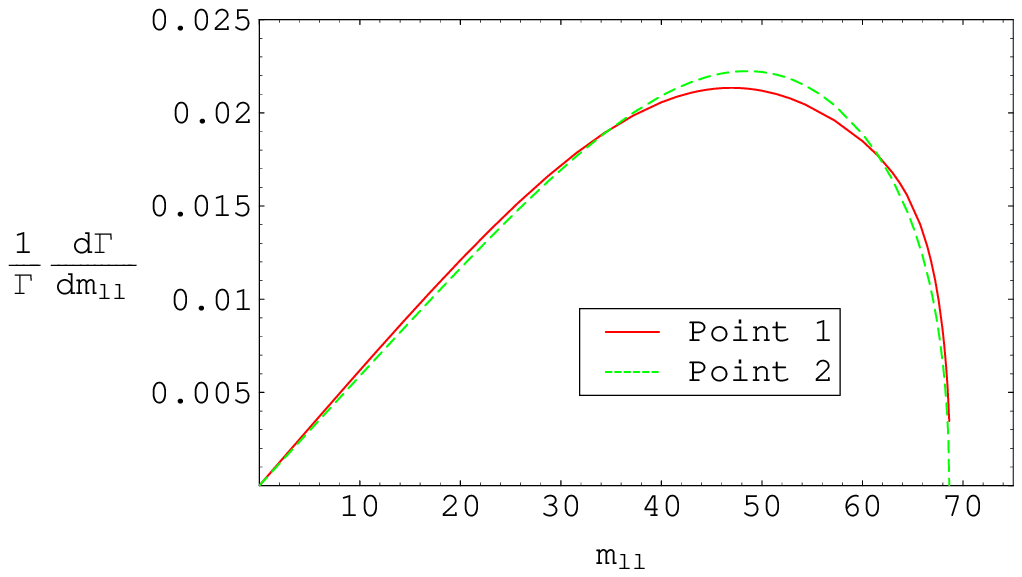}}
    \scalebox{1.28}{\includegraphics{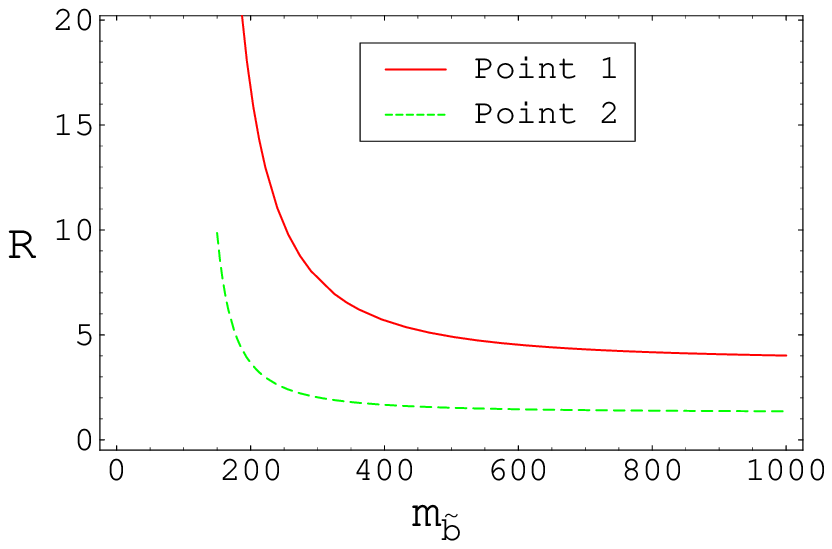}}
  \end{center}
    \caption{While the lepton invariant mass distribution may be degenerate in some cases, $R$ provides another measurement to separate the two.  {\bf Point 1}: $m_{\tilde{l}_L} = 250$ GeV, $m_{\tilde{l}_R} = 500$ GeV, $M_1 = 57$ GeV, $M_2 = 140$ GeV, $\mu = 310$ GeV, and $\tan{\beta} = 4$.  {\bf Point 2}: $m_{\tilde{l}_L} = 255$ GeV, $m_{\tilde{l}_R} = 255$ GeV, $M_1 = 70$ GeV, $M_2 = 140$ GeV, $\mu = -300$ GeV, and $\tan{\beta} = 4$. \label{fig:Rdiff}}
\end{figure}

The reverse is also true: $R$ is a valuable complement to the invariant lepton mass distribution.  Points in SUSY parameter space with very different neutralino mass mixing parameters and left and right slepton masses can conspire to give similar shapes for the normalized invariant mass distribution.  When fitting the SUSY parameters, $R$ could help lift the degeneracy between these two identical lepton invariant mass distributions.  As an example, consider the two nearly identical mass distributions in the top panel of Figure~\ref{fig:Rdiff}.  These two points have very different values of $R$ for most values of the b-squark mass (see two curves in bottom panel).  Taking into account the detector efficiencies (see next section), it should be possible to distinguish the two curves.  This indicates that $R$ has power to help resolve degeneracies left by the lepton invariant mass distribution.

As seen in Figures~\ref{fig:Ratiopt1graphs} and~\ref{fig:Rdiff}, with good knowledge of the neutralino mass mixing parameters and either the slepton or b-squark masses, $R$ can be used to make a measurement of the mass of the b-squark or slepton, respectively.  
In practice, information on the neutralino mass matrix is likely to be incomplete, and $R$ will represent a constraint in the high-dimensional parameter space of supersymmetry breaking parameters.

%%%%%%%%%%Measuring R
\section{Measuring $R$ at the LHC}
Having discussed the utility of $R$, we now make a preliminary assessment of the prospects for measuring it at the LHC.  Of course, both its measurement and interpretation will be dependent on the details of the superpartner spectrum.  We will address these dependencies where appropriate.

\begin{figure}
\begin{center}
\scalebox{0.6}{
  \begin{picture}(315,226) (15,-14)
    \SetWidth{0.5}
    \SetColor{Black}
    \ArrowLine(15,181)(135,181)
    \Gluon(15,181)(135,181){7.5}{7.14}
    \ArrowLine(135,181)(225,211)
    \DashArrowLine(135,181)(165,106){10}
    \ArrowLine(165,106)(225,61)
    \Photon(165,106)(225,61){7.5}{4}
    \Vertex(225,61){2.83}
    \ArrowLine(225,61)(285,16)
    \ArrowLine(225,61)(285,106)
    \Photon(225,61)(285,16){7.5}{4}
    \Text(15,151)[lb]{\Large{\Black{$\tilde{g}$}}}
    \ArrowLine(240,136)(165,106)
    \ArrowLine(300,61)(225,61)
    \Text(310,46)[lb]{\Large{\Black{$\overline{b}, e^+, \mu^+$}}}
    \Text(285,91)[lb]{\Large{\Black{$b, e^-, \mu^-$}}}
    \Text(285,-14)[lb]{\Large{\Black{$\tilde{\chi}_1^0$}}}
    \Text(170,56)[lb]{\Large{\Black{$\tilde{\chi}_2^0$}}}
    \Text(240,146)[lb]{\Large{\Black{$\overline{q}, \overline{b}$}}}
    \Text(225,196)[lb]{\Large{\Black{$q,b$}}}
    \Text(125,136)[lb]{\Large{\Black{$\tilde{q}, \tilde{b}$}}}
  \end{picture}
}
\end{center}
\caption{The gluino decay chain used to measure $R$. \label{fig:gluinodecaychain}}
\end{figure}
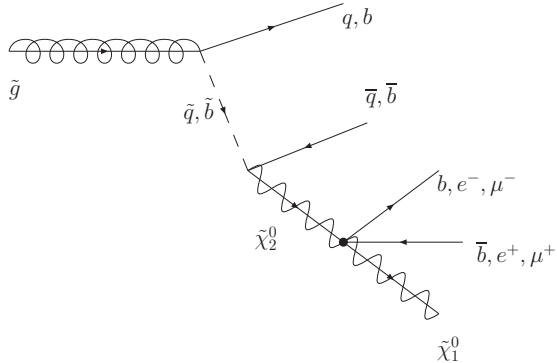

It is likely that main source of $\tilde{\chi}_2^0$ will be gluino cascade decays.  The full cascade decay of the gluino that we consider is shown in Figure~\ref{fig:gluinodecaychain} and was first considered in~\cite{Baer:1987} and ~\cite{Barnett:1988}. 
We make the common assumption $m_{\tilde{\chi}_1^0},m_{\tilde{\chi}_2^0} < m_{\tilde{g}}$.  As alluded to earlier, we require $m_{\tilde{\chi}_2^0} - m_{\tilde{\chi}_1^0} < m_Z$ so that $\tilde{\chi}_2^0$ will undergo three-body decays.  We assume $m_{\tilde{g}} << m_{\tilde{q}}$, ensuring that squark pair production and associated production are sub-dominant to gluino pair production.  This also ensures that $\tilde{g}$ will undergo three-body decays.  Finally, we take $m_{\tilde{g}} < m_{\tilde{\chi}_{3,4}^0}, m_{\tilde{\chi}_2^+}$, forbidding decays of the type $\tilde{g} \rightarrow \tilde{\chi}_{3,4}^0 X$ and $\tilde{g} \rightarrow \tilde{\chi}_{2}^+ X$.  We take the two lightest neutralinos to be mostly gaugino, i,e., $M_1, M_2 << \mu$.  As we emphasize in the next section, with this type of spectrum the events that contribute to $R$, i.e., $\tilde{\chi}_{2}^{0}$ decays, can be isolated readily. 

With the above spectrum, $BR(\tilde{\chi}_2^0 \rightarrow l^+ l^- \tilde{\chi}_1^0)$ is rather small, say, a few percent.  The measurement of $R$ will likely be statistics limited, and a significant measurement will be difficult for $m_{\tilde{g}} \gsim 500$ GeV.  Where specific numbers are cited, we assume $100 \, fb^{-1}$ of integrated luminosity and use $m_{\tilde{g}} = 303$ GeV, $M_1 = 70$ GeV, $M_2 = 140$ GeV, $\mu = 300$ GeV, and $\tan{\beta} = 4$.  Generalizations of this spectra are discussed in the next section.   

  In this section we will consider two different spectra differing in the relation of the mass of the b-squark to the light squarks (denoted collectively as $\tilde{q}$).  We will let $q$ represent the light quarks and explicitly label $b$'s where applicable. 

%%%%%%%%%%%%%%Case 1: mbsquark > mlightsquarks
\subsection{Case 1: $m_{\tilde{b}} \gg m_{\tilde{q}}$}
\label{sec:lightsquarks}
In this case gluinos decay dominantly into light quarks, so the primary source of b-quarks in supersymmetric events will be from the neutralino decays.  A small number of b-squarks are produced via the sub-dominant decay mode of the gluino, but since the gluino branching ratios are proportional to $1/m_{\tilde{b}}^4$ versus $1/m_{\tilde{q}}^4$ we will assume these are negligible\footnote{Even if non-negligible, this source of $b$-jets can be at least partially distinguished from $b$-jets from neutralino decays with additional kinematic cuts, e.g. $m_{bb}$.}.  With these assumptions, we count the ratio of two different types of decays: $\tilde{g} \rightarrow q \overline{q} \tilde{\chi}_2^0 \rightarrow q \overline{q} b \overline{b} \tilde{\chi}_1^0$ versus $\tilde{g} \rightarrow q \overline{q} \tilde{\chi}_2^0 \rightarrow q \overline{q} l^+ l^- \tilde{\chi}_1^0$, where $l^+$ is either $e^+$ or $\mu^+$, i.e. we measure the ratio 
\begin{equation}
\frac{\# (4\,\textrm{jets}) (2\,\textrm{b-jets}) l^+ l^- + \slashed{E}_t}{\# (4\,\textrm{jets}) l^+ l^- l^{'+} l^{'-} + \slashed{E}_t},
\end{equation}
subject to some additional kinematic cuts to be discussed below.  Under our assumptions about the spectrum, after correcting for detector acceptances and efficiencies, this ratio yields a measurement of $R$.  To get a feel for the number of events available for the measurement, we used PYTHIA~\cite{PYTHIA} to calculate the cross section and branching ratios and the PGS4 detector simulation~\cite{PGS4} to estimate the detector efficiency.  It is necessary to apply a set of cuts to eliminate Standard Model backgrounds, which we now discuss.  

We simulated 10000 events with the topology $\tilde{g} \tilde{g} \rightarrow q \bar{q} q \bar{q} + 2 \chi_{2} \rightarrow q \bar{q} q \bar{q}+ 4l+ 2 \chi_{1}$.  Using PGS, the efficiency for these events for the signal $l^+ l^- l'^+ l'^-$ ( $p_{t}(l)  > 15$ GeV) and 4 jets ($p_{t} > 15$ GeV) with $\slashed{E}_t > 100$ GeV is estimated at $\approx 10\%$.  Folding in the gluino production cross section and branching ratios, the expected number of signal events for our sample spectrum with squark masses $m_{\tilde{q}} = 600$ GeV and $(m_{\tilde{b}}, m_{\tilde{l}}) = (1 \textrm{ TeV}, 1 \textrm{ TeV})$ is 732 and for $(m_{\tilde{b}}, m_{\tilde{l}}) = (1 \textrm{ TeV}, 400 \textrm{ GeV})$ is 63 with $m_{\tilde{g}} = 303$ GeV. This sharp decrease is precisely due to the destructive interference of the slepton discussed earlier, which decreases $BR(\tilde{\chi}_2^0 \to l^+ l^- \tilde{\chi}_1^0)$.  The dominant SM physics background for this signal would be $t\overline{t}\, + (>2 \textrm{ jets})$, with both tops decaying leptonically and each resultant b decaying semi-leptonically.  However, after isolation cuts on the leptons this background is small.  Other backgrounds (such as $ZZ$ production), are largely eliminated by imposing the missing energy cut.

For the final state $l^+ l^- b \overline{b} \, + $ 4 jets $+\,\slashed{E}_t$, we need make more stringent cuts to reduce the background.  We use the set of cuts and related backgrounds quoted in~\cite{Hinchliffe:1999}:  
\begin{itemize}
\item A pair of isolated electrons or muons of opposite charge and the same flavor with $p_{t}(l) > 10$ GeV and $|\eta| < 2.5$;
\item The effective mass, formed from the four hardest jets and the missing energy, $M_{eff} \equiv p_{t,1} + p_{t,2} + p_{t,3} + p_{t,4} + \slashed{E}_t > 500$ GeV;
\item $\geq$ 4 jets with $p_t > 50$ GeV and $p_{t,1} > 100$ GeV;
\item Transverse sphericity, $S_t > 0.2$;
\item $\slashed{E}_t > 0.2 M_{eff}$;
\item $m_{ll} < 90$ GeV.  (Added)
\end{itemize}

With these cuts, the backgrounds (conservatively estimated from the Figure 11 in \cite{Hinchliffe:1999}) are $\approx 2800$ events per $10 \, fb^{-1}$.  Applying these cuts to the gluino cascade signal yields an efficiency of $7.5\%$,  so the expected signal is $\approx 7400$ events per $10 \, fb^{-1}$ for the sample spectrum.  The signal should be easily observable above background.  This estimate is prior to placing any requirements on the number of b-jets.  We must require at least one b-tag to ensure that our neutralino is decaying as $\tilde{\chi}_2^{0} \rightarrow \tilde{\chi}^{0}_{1} b \bar{b}$.  This will decrease the signal, but we would expect an even larger reduction of SM background.  After all, the background includes SM processes that do not include any b-jets.  We also expect that some of the SM backgrounds, for instance $t \overline{t}jjjj \rightarrow (l^+ \nu b) (l^{'-} \overline{\nu}^{'} \overline{b}) jjjj$, to be able to be lessened by using flavor subtraction methods.   In principle, this might allow a softening of the above cuts.  With an extra requirement of one b-tag, the PGS4 calculated efficiency is $\approx 3.2 \%$.  The expected number of signal events for our sample spectrum with $(m_{\tilde{b}}, m_{\tilde{l}}) = (1 \textrm{ TeV}, 1 \textrm{ TeV})$ is 1334 and for $(m_{\tilde{b}}, m_{\tilde{l}}) = (1 \textrm{ TeV}, 400 \textrm{ GeV})$ is 98.  It should be emphasized that the cuts here represent an existence proof for this measurement.  While these cuts are well known to be effective at isolating SUSY events, they are not completely optimized for the spectrum considered here.  Once more information is known about the scale of supersymmetry, cuts should tailored to maximize the signal to background ratio, and to minimuze the potential pollution of the signal by other SUSY events.  

\begin{figure}
  \begin{center}
    \scalebox{1.5}{\includegraphics{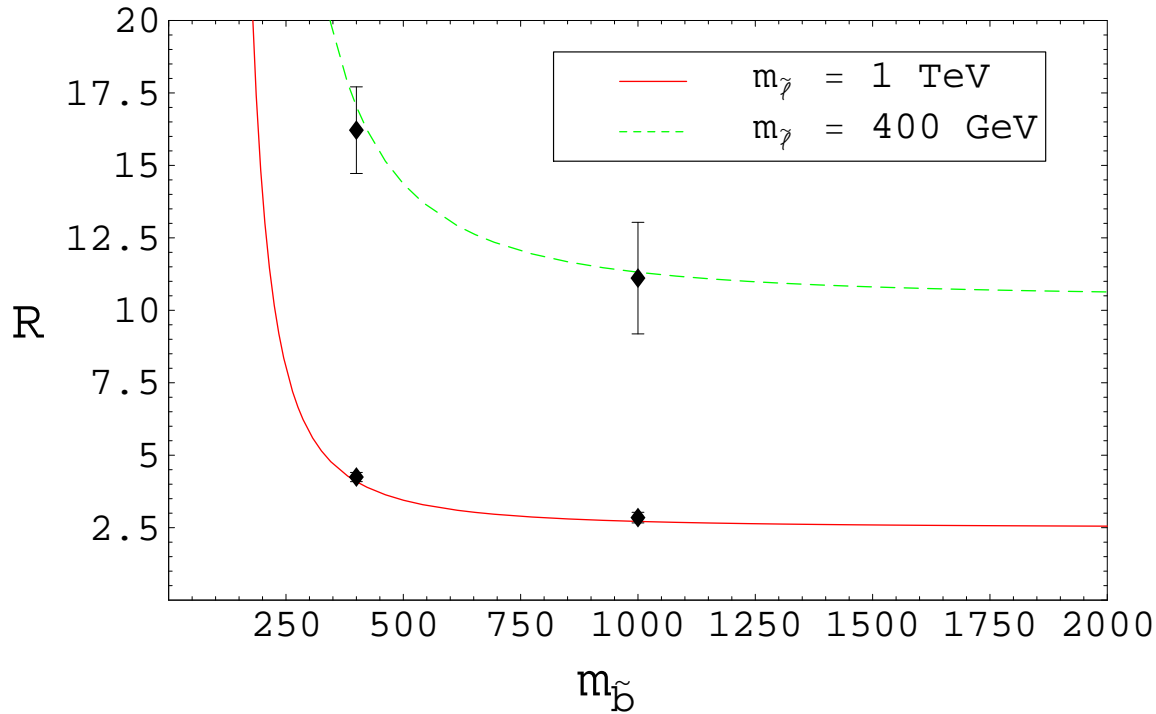}}
  \end{center}
    \caption{Sample $R$ measurements with expected error bars as a function of the b-squark mass for the sample points with $m_{\tilde{l}} = 1$ TeV and $m_{\tilde{l}} = 400$ GeV.  The point at $m_{\tilde{b}} = 1$ TeV uses the method and efficiencies of Case 1 and the point at $m_{\tilde{b}} = 400$ GeV uses the method and efficiencies of Case 2.
  \label{fig:samplept70graphs}}
\end{figure}

After applying the two sets of above cuts for the two classes of signal events, for $100 \, fb^{-1}$ of data we can reconstruct 
$R = 2.8 \pm 0.2$ for the point at $(m_{\tilde{b}}, m_{\tilde{l}}) = (1 \textrm{ TeV}, 1 \textrm{ TeV})$ and $R = 11.1 \pm 1.9$ for the point at $(m_{\tilde{b}}, m_{\tilde{l}}) = (1 \textrm{ TeV}, 400 \textrm{ GeV})$ seen in Figure~\ref{fig:samplept70graphs}.   The error bars are statistical.  In practice, it would be important to understand the acceptances for the cuts in detail. 

%%%%%%%%%%%%%%Case 2: mbsquark < mlightsquarks
\subsection{Case 2: $m_{\tilde{b}} < m_{\tilde{q}}$}
In this case, the gluinos dominantly decay to final states with two b-quarks.  The fact that the SUSY events are so b-rich can be used to enhance the signal over the SM background.  However, the extra b-jets also complicate the interpretation of supersymmetric events:  there is an ambiguity as to whether a given b-jet comes from gluino decay or from neutralino decay. This can be avoided if we measure 
\begin{equation}
\frac{\# (6\,\textrm{jets} (\geq 5\,\textrm{b-jets})) l^+ l^- + \slashed{E}_t}{\# (4\,\textrm{jet} (\geq 3\, \textrm{b-jets})) l^+ l^- l^{'+} l^{'-} + \slashed{E}_t}.
\end{equation}
All SM backgrounds are negligible.

Unfortunately, the efficiency for tagging at least five b-jets is rather small. Because of the b-rich sample, however,  we can loosen the cuts on jet $p_t$ and $\slashed{E}_t$ and still avoid standard model backgrounds.  Requiring the leptons be opposite sign and same flavor, at least six jets ($p_t > 20$ GeV) with at least five b-tags, and $\slashed{E}_t > 30$ GeV, we estimate an efficiency of $\approx 0.8 \%$.  For our chosen point with $m_{\tilde{q}}= 1$ TeV, for $(m_{\tilde{b}}, m_{\tilde{l}}) = (400 \textrm{ GeV}, 1 \textrm{ TeV})$ we get an expected signal of 2416 events and for $(m_{\tilde{b}}, m_{\tilde{l}}) = (400 \textrm{ GeV}, 400 \textrm{ GeV})$ we get an expected signal of 821 events after $100 \, fb^{-1}$ of integrated luminosity.  
For the four-lepton signal we remove the requirement of any b-tags.  This will not cause any confusion as long as $BR(\tilde{g} \to q \bar{q} \tilde{\chi}_2^0)$ is small.  
We thus require four jets and all four leptons to have $p_t > 15$ GeV and $\slashed{E}_t > 100$ GeV just as in case 1.  For an estimated efficiency of $10\%$,  we get 3427 expected events for  $(m_{\tilde{b}}, m_{\tilde{l}}) = (400 \textrm{ GeV}, 1 \textrm{ TeV})$ and 305 events for  $(m_{\tilde{b}}, m_{\tilde{l}}) = (400 \textrm{ GeV}, 400 \textrm{ GeV})$.  
The measured value for $100 \, fb^{-1}$ of data is $R = 4.2 \pm 0.2$ for the point at $(m_{\tilde{b}}, m_{\tilde{l}}) = (400 \textrm{ GeV}, 1000 \textrm{ GeV})$ and $R = 16.2 \pm 1.5$ for the point at $(m_{\tilde{b}}, m_{\tilde{l}}) = (400 \textrm{ GeV}, 400 \textrm{ GeV})$, seen in Figure~\ref{fig:samplept70graphs}. Again, error bars are statistical only.

To make a valid measurement of $R$, gluino decays that give final states identical to $\tilde{g} \rightarrow  b \overline{b} \tilde{\chi}_2^0 \to b\overline{b} l^+ l^- \tilde{\chi}_1^0$ must be suppressed.  Because in this case the third generation squarks are light, one worry is the decay chain $\tilde{g} \rightarrow  b \overline{t} \tilde{\chi}_1^+ \to b \overline{b} l^{'+} l^- \nu^{'} \bar{\nu} \tilde{\chi}_1^0$.  For a small enough gluino mass, decays to charginos are kinematically forbidden due to the large top mass, potentially leading to a larger branching ratio for $\tilde{g} \rightarrow \chi_{2}^{0} b \bar{b}$ and no pollution from the chargino decays.  For larger gluino masses, however, gluino decays to charginos can be competitive with the decays to neutralinos.  After all, in the limit where the $\chi_{2}^{0}$ is pure wino, the gluino branching ratio $\tilde{g} \rightarrow  b \overline{b} \tilde{\chi}_2^0$ can be related to  $\tilde{g} \rightarrow  t \overline{b} \tilde{\chi}_2^+$ by a simple Clebsch-Gordon coefficient.  Requiring $m_{ll} < m_Z$ will reduce this background somewhat.  Perhaps more importantly, the leptons coming from the chargino decay and the top decay are not necessarily the same flavor, so flavor subtraction may be possible.  In particular, as long as $BR(t \to b l^+ \nu) \times BR(\tilde{\chi}_1^+ \to l^+ \nu \tilde{\chi}_1^0)$ is not much greater than $BR(\tilde{\chi}_2^0 \to l^+ l^- \tilde{\chi}_1^0)$, we expect to be able to perform this subtraction, so $R$ can still be measured accurately.  This will be the case over most of parameter space.

%%%%%%%%%%Other spectra where R can be measured.
\section{Generalizing the spectra}
In the previous section, we chose very specific benchmark to discuss efficiencies and event rates.   However, the measurement of $R$ can be generalized to other spectra.  In this section, we discuss the cases in which interpretation of $R$ remains straightforward.  A non-negotiable requirement on the spectrum is $m_{\tilde{\chi}_2^0} - m_{\tilde{\chi}_1^0} < m_Z$, i.e. that the neutralino still undergo three-body decays.  This requirement must be kept for the analysis of section 2 to hold.  

Since the b-squark only constructively interferes in neutralino decays, lowering the b-squark mass will raise $BR(\tilde{\chi}_2^0 \rightarrow b \overline{b} \tilde{\chi}_1^0)$ and hence lower $BR(\tilde{\chi}_2^0 \rightarrow l^+ l^- \tilde{\chi}_1^0)$.  For the point considered in this paper, the number of four lepton events is small, and the statistical error associated with this sample introduces the the largest source of error in R.  Thus, the enhancement of the branching ratio to $b$ quarks will degrade the $R$ measurement.  If $m_{\tilde{b}} < m_{\tilde{g}}$, then on-shell gluino decays to b-squarks are allowed.  On-shell b-squarks will not affect the accuracy of our measurement so long as $BR(\tilde{b} \rightarrow \tilde{\chi}_2^0 b)$ is still large.  In fact, the same argument applies to any on-shell gluino decays to any of the squarks.  On-shell decays of gluinos may help in determining the gluino mass and b-squark mass, see discussion of Point 3 in~\cite{Hinchliffe:1996}.

If the squarks are not very heavy,  associated production of $\tilde{g} \tilde{q}$ will contribute a large source of SUSY events at the LHC (and can even be the dominate production mechanism).  How will the presence of these events affect the measurement of R?  Assuming the squarks are heavier than the gluino, squarks will likely decay primarily to gluinos.  The final state associated with these events would be identical to gluino pair production, but with the presence of an extra jet.  Understanding differences in acceptances for this class of events would be important, but would not otherwise have a significant effect on the measurement of $R$ as presently defined.   If the squarks become significantly lighter than the gluino, their pair production could become the dominant source of SUSY events at the LHC.  Furthermore, they can then have significant branching fractions to charginos and neutralinos.  Decays $\tilde{q} \rightarrow q \tilde{\chi}_2^0$ could conceivably be used in a manner analogous to the one described in this paper. However, the optimal set of cuts used to distinguish events from Standard Model background (see section \ref{sec:lightsquarks}) would look very different.  

If $m_{\tilde{\chi}_{3,4}^0} < m_{\tilde{g}}$, then all of the neutralinos will appear in gluino cascade decays.  Decays such as $\tilde{\chi}_{3,4}^0 \rightarrow l^+ l^- \tilde{\chi}_1^0$ could, in principle, pollute the signal. If the assumption  $M_1, M_2 << \mu$ still remains, the heavier neutralinos are primarily Higgsino and are much heavier than $\tilde{\chi}_2^0$.  Decays of gluinos to $\tilde{\chi}_{3,4}^0$ may be suppressed relative to $\tilde{\chi}_2^0$ by small Yukawa couplings and the larger mass of the neutralinos.  Furthermore, the heavy neutralinos will likely decay primarily via the two-body decays, into e.g., a lighter neutralino and a $Z^0$.  Thus, cuts on the invariant mass of the lepton or b-pairs should reduce this potential pollution of the $R$ measurement.

%%%%%%%%%%Conclusion
\section{Conclusion}
In this study, we formulated an observable $R$ sensitive to the interference of the slepton and b-squark diagrams in neutralino decays.  It complements the work of \cite{Nojiri:1999,Birkedal:2005} on lepton invariant mass distributions in neutralino decay.  Measuring both quantities can help lift some of the degeneracies found in different parts of SUSY parameters space.

In principle, combining information from the lepton invariance mass distribution with the invariant mass distribution of the $b$ quarks coming from decays of the type $\chi_{2}^{0} \rightarrow \chi_{1}^{0} b \bar{b}$ contains information similar to the variable described here.  However, the efficacy of this measurement will be dependent on the ability of the detectors to reconstruct the invariant masses of the $b$ jets.  Whether this can be done at a level that allows the extraction of any information beyond that discussed here is an area for future investigation.

It should also be noted that events where there are two $\chi_{2}^{0}$ decays are quite distinctive.  So, in principle, they might have uses beyond a measurement of $R$.  For example, they can be used to make a rather unambiguous measurement of different branching ratios of the gluinos by determining the ratio:
\begin{equation}
R_{\tilde{g}} = \frac{BR(\tilde{g} \rightarrow b \overline{b} \tilde{\chi}_2^0)}{BR(\tilde{g} \rightarrow q \overline{q} \tilde{\chi}_2^0)}
\end{equation}
by looking at events with four leptons.  The requirement of four leptons ensures that we are looking at decays $\tilde{\chi}_2^0 \rightarrow l^+l^- \tilde{\chi}_1^0$.  $R_{\tilde{g}} << 1$ corresponds to case 1 of section 3 and $R_{\tilde{g}} >>1$ corresponds to case 2.  This measurement would suffer from the small branching ratio of neutralinos to decays to leptons, but is quite clean and would be complementary to other approaches. Knowledge of this branching ratios gives a constraint on the squark masses and squark couplings to neutralinos.  

%%%%%%%%%%%Acknowledgements
\section{Acknowledgments}  

The authors would like to thank Dan Amidei, Can Kilic, Jesse Thaler, Lian-Tao Wang, and James Wells for helpful discussions and comments.  The work of AP and DJP is supported by the Michigan Center for Theoretical Physics. DJP also receives support from a Ford Foundation Fellowship.

%%%%%%%%%%%Appendix: Calculating 3-Body Decay Amplitudes
\appendix
\section{Calculating 3-Body Decay Amplitudes}
The neutralino three-body partial width was first calculated in~\cite{Bartl:1986}, however here we will closely follow the work and notation of~\cite{Nojiri:1999}.  We first start by looking at the mass matrix of the neutralinos in the $\{\tilde{B},\tilde{W},\tilde{H_1},\tilde{H_2}\}$ basis:
\begin{equation}
  \mathbf{M}_N = \left(\begin{array}{cccc}
  M_1 & 0 & -m_Z \cos \beta \sin \theta_W & m_Z \sin \beta \sin \theta_W \vspace{.2cm}\\
  0 & M_2 & m_Z \cos \beta \cos \theta_W & -m_Z \sin \beta \cos \theta_W \vspace{.2cm}\\
  -m_Z \cos \beta \sin \theta_W & m_Z \cos \beta \cos \theta_W & 0 & -\mu \vspace{.2cm}\\
  m_Z \sin \beta \sin \theta_W & -m_Z \sin \beta \cos \theta_W & -\mu & 0 \end{array} \right).
  \label{eq:NeutralinoMassMatrix}
\end{equation}
We define the mass eigenstates as: 
\begin{equation}
\tilde{\chi}_i^0=N_{i1}\tilde{B}+N_{i2}\tilde{W}+N_{i3}\tilde{H_1}+N_{i4}\tilde{H_2},
\end{equation}
where $N_{ij}$ is the mixing matrix.  We neglect the possibility of CP violation and take the mass mixing matrix to be real.  We then get a Lagrangian 
\begin{equation}
  \mathcal{L}_{int} = -\frac{g}{\sqrt{2}} \overline{\tilde{\chi}_A^0}(a_{AX}^f P_L +b_{AX}^f P_R)f \tilde{f}_X^{\dagger} + h.c. 
+ \frac{g_Z}{2} z_{BA}^{(\tilde{\chi}^0)} \overline{\tilde{\chi}_B^0} \gamma^{\mu} \gamma_5 \tilde{\chi}_A^0 Z_{\mu}^0,
\end{equation}
where $g_Z = \sqrt{g^{\prime 2} + g^2}$, and
\begin{equation}
  z_{BA}^{(\tilde{\chi}^0)} = N_{B3} N_{A3} - N_{B4} N_{A4};
\end{equation}
\begin{equation}
  a^f_{AX} = 2(T_3 N_{A2} + Y_L N_{A1} \tan{\theta_W})\delta_{X,L} + \frac{m_f}{m_W \cos{\beta}} N_{A3} \delta_{X,R};
\end{equation}
\begin{equation}
  b^f_{AX} = -2 Y_R N_{A1} \tan{\theta_W} \delta_{X,R} + \frac{m_f}{m_W \cos{\beta}} N_{A3} \delta_{X,L}.
\end{equation}

Then the squared, spin-averaged amplitude for $\tilde{\chi}_A^0 (p) \longrightarrow \tilde{\chi}_B^0(\overline{p}) f(q) \overline{f}(\overline{q})$  coming from the two diagrams shown in Figure~\ref{fig:neutrfeyn} is
\begin{eqnarray}
  \left| \mathcal{M} \right|^2 &=& 2(A_{LL}^2 +A_{RR}^2)(m_{\tilde{\chi}_A^0}^2-y)(y-m^2_{\tilde{\chi}_B^0}) 
   + 2(A_{LR}^2 +A_{RL}^2)(m_{\tilde{\chi}_A^0}^2-x)(x-m^2_{\tilde{\chi}_B^0})\nonumber \\ 
   &-& 4(A_{LL} A_{RL} + A_{RR} A_{LR}) m_{\tilde{\chi}_A^0} m_{\tilde{\chi}_B^0} z,
\end{eqnarray}
with
\begin{eqnarray}
  A_{LL} &=& \frac{g_Z^2}{2} \frac{ z_{BA}^{\tilde{\chi}^0} (T_3 - Q_f \sin^2{\theta_W}) }{z-m_Z^2} - \frac{g^2}{2} \frac{a_{AL}^f a_{BL}^f}{y-m^2_{\tilde{f}_L}}- \frac{g^2}{2} \frac{a_{AR}^f a_{BR}^f}{y-m^2_{\tilde{f}_R}} ;\nonumber \\
  A_{RL}&=&-A_{LL}(y \leftrightarrow x) ;\nonumber \\
  A_{LR} &=& \frac{g_Z^2}{2} \frac{ z_{BA}^{\tilde{\chi}^0} (- Q_f \sin{\theta_W}^2) }{z-m_Z^2} - \frac{g^2}{2} \frac{b_{AL}^f b_{BL}^f}{y-m^2_{\tilde{f}_L}}- \frac{g^2}{2} \frac{b_{AR}^f b_{BR}^f}{y-m^2_{\tilde{f}_R}}; \nonumber \\
  A_{RR}&=&-A_{LR}(y \leftrightarrow x). \label{Matrixelem}
\end{eqnarray}
Here $x$, $y$, and $z$ are the invariant masses formed by
\begin{equation}
  x = (\overline{p}+q)^2,\, y=(\overline{p}+\overline{q})^2 ,\, z = (q+\overline{q})^2.
\end{equation}
The derivative of partial decay width with respect to the invariant mass of the fermion pair is then~\cite{PDG}.
\begin{equation}
\frac{d\Gamma}{dm_{f\overline{f}}}(\tilde{\chi}_A^0 \longrightarrow \tilde{\chi}_B^0 f \overline{f}) = 
  \frac{1}{(2\pi)^3} \frac{1}{4 m^2_{\tilde{\chi}_A^0}}\left| \mathcal{M} \right|^2 \left|\vec{\overline{p}}\right| \left| \vec{q} \right|
\end{equation}
where
\begin{equation}
 \left|\vec{\overline{p}}\right| = \frac{(m^2_{f\overline{f}}-4 m^2_f)^{1/2} m_{f\overline{f}} }{2 m_{f\overline{f}} }, 
 \left| \vec{q} \right| = \frac{[(m^2_{\tilde{\chi}_A^0}-(m_{f\overline{f}}+m_{\tilde{\chi}_B^0})^2 )
             (m^2_{\tilde{\chi}_A^0}-(m_{f\overline{f}}-m_{\tilde{\chi}_B^0})^2 )]^{1/2}}{2 m_{\tilde{\chi}_A^0} }.
\end{equation}
Finally, integration over the invariant mass from $2 m_f$ to $m_{\tilde{\chi}_2^0} - m_{\tilde{\chi}_1^0}$ gives the partial width.

We did not include the contribution from the Higgs sector in the above calculation.  This will be most important when the neutralinos are extremely gaugino-like since the coupling through the $Z^0$ will be suppressed. For all the parameters considered here, the contribution from including in the partial width due to Higgs boson exchange is less than  $5 \%$.  For small or moderate values of $\tan{\beta}$ we can ignore the Higgs boson contribution due to the smallness of the bottom Yukawa coupling.  Areas of parameter space where Higgs boson exchange does become important are discussed in \cite{Bartl:1999}.

%%%%%%%%%%%Bibliography


\begin{thebibliography}{99}

\bibitem{ATLAStdr}
  ATLAS Collaboration,
  ``ATLAS detector and physics performance. Technical design report.  Vol. 2,''
  CERN-LHCC-99-15 (1999).

\bibitem{Hinchliffe:1996}
  I. Hinchliffe, F.E. Paige, M.D. Shapiro, J. S\"oderqvist, and W. Yao,
  ``Precision SUSY Measurements at LHC'',
  Phys. Rev. D 55, 5520 (1997)
  [arXiv:hep-ph/9610544]

\bibitem{Nojiri:1999}
  Mihoko M. Nojiri and Youichi Yamada,
  ``Neutralino decays at the Cern LHC'',
  Phys. Rev. D 60:015006, 1999
  [arXiv:hep-ph/9902201].

\bibitem{Birkedal:2005}
  A. Birkedal, R.C. Group, and K. Matchev,
  ``Slepton Mass Measurements at the LHC'',
  In the Proceedings of 2005 International Linear Collider Workshop (LCWS, 2005), Stanford, California, 18-22 Mar 2005, pp 0210.
  [arXiv:hep-ph/0507002]
%5
\bibitem{Baer:1998}
  H. Baer, C. Chen, M. Drees, F. Paige, and X. Tata,
  ``Probing Minimal Supergravity at the Cern LHC Large $\tan{\beta}$'',
  Phys. Rev. D 59:055014, 1999
  [arXiv:hep-ph/9809223]

\bibitem{Baer:1995}
  H. Baer, Chih-hao Chen, Chung Kao, and Xerxes Tata,
  ``Supersymmetry reach of an upgraded Tevatron collider'',
  Phys. Rev. D 52, 1565 (1995)

\bibitem{Mrenna:1996}
  S. Mrenna, G. L. Kane, G. D. Kribs, and J. D. Wells,
  ``Possible signals of constrained supersymmetry at a high luminosity Fermilab Tevatron Collider'',
  Phys. Rev. D 53, 1168 (1996)

\bibitem{Bartl:1999}
  A. Bartle, W. Majerotto, and W. Porod,
  ``Large Higgs Boson Exchange Contribution in Three-Body Neutralino Decays,'',
  Phys. Lett. B 465:187-192, 1999
  [arXiv:hep-ph/9907377]

\bibitem{Arnowitt:1995}
  R. Arnowitt and P. Nath,
  ``Predictions of Neutralino Dark matter event rates in minimal supergravity unification'',
  Phys. Rev. D54:2374-2384(1996),
  [arXiv:hep-ph/9509260]
%10
\bibitem{Baer:1987}
  Howard Baer, V. Barger, Debra Karatas, and Xerxes Tata,
  ``Detecting Gluinos at hadron colliders'',
  Phys. Rev. D 36, 96 (1987)

\bibitem{Barnett:1988}
  R. Michael Barnett, John F. Gunion, and Howard H. Haber,
  ``Gluino decay patterns and signatures'',
  Phys. Rev. D 37 1892 (1988)

\bibitem{PYTHIA}
  Torb\"orn Sj\"orstrand, Stephen Mrenna, and Peter Skands,
  ``PYTHIA 6.4 Physics and Manual'',
  JHEP 0605:026, 2006
  [arXiv:hep-ph/0603175]

\bibitem{PGS4}
  John Conway,
  PGS4 detector simulation,  
  http://www.physics.ucdavis.edu/$\sim$conway/research/software/pgs/pgs4-general.htm

\bibitem{Hinchliffe:1999}
  I. Hinchliffe and F. Paige,
  ``Measurements in SUGRA Models with Large $\tan{\beta}$ at the LHC''
  Phys. Rev. D 61:095011, 2000
  [arXiv:hep-ph/9907519]
%15
\bibitem{Bartl:1986}
  A. Bartl, H. Fraas, W. Majerotto,
  ``Production and decay of neutralinos in $e^+ e^-$ annihilation'',
  Nucl. Phys. B278 (1986) 1

\bibitem{PDG}
 W.-M. Yao et. al.,
 ``Review of Particle Physics, Particle Data Group'',
 Journal of Physics G 33, 1 (2006)

\end{thebibliography}
\end{document}